\documentclass[footinbib,twocolumn,showpacs,amsmath,amstex,amssymb,mathfonts,superscriptaddress,prl]{revtex4}
\usepackage{graphicx}
\usepackage{color}
\usepackage{bm}

\usepackage{graphicx}
\usepackage{color}
\usepackage{bm}
\usepackage{amsmath}
\usepackage{amssymb}
\usepackage{amsthm}
\usepackage{amsfonts}
\usepackage{multirow}
\usepackage{makecell}
\usepackage{float}
\usepackage{comment}
\usepackage{enumitem}

\usepackage{bbm}

\begin{document}

\title{Emergent Commensurability from Hilbert Space Truncation in Fractional Quantum Hall Fluids}
\author{Bo Yang} 
\affiliation{Division of Physics and Applied Physics, Nanyang Technological University, Singapore 637371.}
\affiliation{Institute of High Performance Computing, A*STAR, Singapore, 138632.}

\date{\today}
\begin{abstract}
We show that model states of fractional quantum Hall fluids at all experimentally detected plateaus can be uniquely determined by imposing translational invariance with a particular scheme of Hilbert space truncation. The truncation is based on classical local exclusion conditions, motivated by constraints on physical measurements. The scheme allows us to identify filling factors, topological shifts and clustering of topological quantum fluids universally without resorting to microscopic Hamiltonians. This prompts us to propose the notion of emergent commensurability as a fundamental property for many known FQH states, which allows us to predict families of new FQH state that can be realised \emph{in principle}. We also discuss the implications of certain missing states proposed from other phenomenological approaches, and suggest that the physics of FQH effect could fundamentally arise from the algebraic structure of the Hilbert space in a single Landau level.
\end{abstract}

\maketitle 

A large number of fractional quantum Hall (FQH) states with distinct topological orders have been observed experimentally and proposed theoretically, ever since the surprising discovery of the quantised Hall conductivity at $1/3$ filling factor\cite{prange,laughlin}. The physics of the FQH effect is mainly derived from the formation of an incompressible quantum fluid with a charge excitation gap, which could be realised at specific rational filling factors when a two-dimensional electron gas system is subject to a perpendicular magnetic field. We now understand that both Abelian and non-Abelian FQH states are likely observed in the experiments\cite{prange,sarmabook}. In addition to the single component FQH states (e.g. the Read-Rezayi series\cite{mr,rr}), there can also be multi-component or hierarchical states from the coexistence of more than one type of quantum fluids in a strongly correlated manner\cite{haldane1,jain1}. 

There has been much development in the microscopic theories of the FQH effect since the first proposition of the Laughlin wavefunctions\cite{laughlin} and later on, the model Hamiltonians\cite{haldane83}. One major approach is the phenomenological formation of ``composite fermions" (CF) with flux attachment\cite{jain1}, and the parton construction inspired from it\cite{jain2,wen1}. It leads to the systematic construction of microscopic wavefunctions for almost all observed and proposed FQH states\cite{jainbook}. Another major approach is to exploit the rich algebraic structures of many-body wavefunctions in a single Landau level (LL) on genus 0 geometries (e.g. sphere or disk), leading to very efficient constructions of microscopic model wavefunctions with the Jack polynomial formalism\cite{jack,bernevig,yang1,yang2}. The method is particularly useful for the Read-Rezayi (RR) series including the coveted non-Abelian states, revealing the particle clustering properties in an intuitive manner. The Jack polynomial formalism and the related techniques are also closely linked to the wavefunction constructions from parafermion correlators in conformal field theory (CFT)\cite{mr,rr}, and in contrast to the CF approach, in many cases model projection Hamiltonians can be found\cite{simon}, of which the constructed wavefunctions are unique zero energy ground states.

From a theoretical point of view, we can characterise the FQH states with the following expression:
\begin{eqnarray}\label{expression}
N_{\phi}=\frac{q}{p}\left(N_e+S_e\right)-S_\phi
\end{eqnarray}
where the system size is given by the number of orbitals $N_\phi$ and the number of electrons $N_e$. In the thermodynamic limit when both $N_\phi,N_e\rightarrow\infty$, the filling factor is given by $\nu=p/q$, and $S_e,S_\phi$ are integer topological shifts for the electrons and orbital\cite{wen2,footnote} respectively. Note that $p,q$ do not have to be co-prime. While the phenomenological CF formalism is extremely useful in conjecturing about possible combinations of $[p,q,S_e,S_\phi]$, it is most successful for Abelian states, and the majority of the CF wavefunctions do not seem to have a local model Hamiltonian\cite{sreejith}. The local projection Hamiltonians (for RR series and beyond) can serve as model Hamiltonians for many Jack polynomials or CFT based wavefunctions, including the non-Abelian ones, but most of them do not have unique zero energy ground states\cite{simon,jack}. The generalisation to Abelian multicomponent FQH states in this picture is also difficult. The key question we ask here is how to determine if an FQH state can form \emph{in principle} at a particular $[p,q,S_e,S_\phi]$. This is because at the fundamental level, these topological indices should not depend on any local operators, including Hamiltonians.

In this paper, we focus on spin-polarised states, and propose a new perspective in the general understanding of the FQH effects. This perspective is based on a number of physically motivated principles and strong numerical evidence, leading to the predictions of many new FQH states with explicit model wavefunctions and topological indices. It also leads to very efficient ways of numerically constructing model wavefunctions, including previously known ones that cannot be written as Jack polynomials and thus could only be obtained by expensive numerical diagonalisation of model Hamiltonians with few-body interactions. 

More specifically, we present a simple set of criteria for the constraints on physical measurements of the quantum Hall fluids, which we term as local exclusion conditions (LECs). They can unambiguously determine $[p,q,S_e,S_\phi]$ for many FQH phases, \emph{without} resorting to microscopic Hamiltonians or other prior conjectures. Interestingly, the LECs are ``classical" constraints on the reduced density matrices of the quantum Hall fluids, but here we use them to determine emergent topological properties that are purely quantum phenomena. This motivates us to propose the concept of intrinsic ``commensurability" for the FQH physics, that is reminiscent of, though not equivalent to, the generalised Pauli exclusion principles\cite{haldane2} and clustering properties\cite{jack}. 

We start by examining some general ingredients for the incompressible FQH states. The ground state of a particular FQH phase needs to be the unique, highest density state that is translationally invariant. Translational invariance is needed so there are no gapless Goldstone modes, and we need the ground state to be of the highest density for it to be incompressible. If such a ground state is not unique, it generally also indicates gapless excitations\cite{oshikawa}, since the degeneracy could be split by small perturbations, unless it is protected by symmetry or topology\cite{footnote1}. We will also assume for any FQH phase, Eq.(\ref{expression}) should be valid for all allowed system sizes, and is not just valid in the asymptotic limit when $N_e$ or $N_\phi$ goes to infinity. 

We also conjecture there exists a minimal description as given by a minimal model ground state living in a truncated Hilbert space, for many (if not all) FQH phases. Here we define the ``minimal" model state to be the state that captures all the ground state topological properties of the topological phase, but made from the smallest possible set of bases, out of all bases permissible by quantum numbers (e.g. momentum) in the Hilbert space. The RR series are explicit examples where part of the translationally invariant Hilbert space in a single LL is truncated, as manifested by their entanglement spectrum\cite{haldane3}. The integer quantum Hall effect also fits in this paradigm: while the permissible Hilbert space includes all Landau levels and realistic ground states have small components in higher LLs, the minimal model ground state is the Vandermonde of the lowest filled LLs with all higher LLs truncated. 

We now show how ``commensurability" of FQH systems arises from these principles, particularly with the important roles played by translational invariance and Hilbert space truncation. We use spherical geometry, so translational invariance is equivalent to rotational invariance, and take the model wavefunction of the Laughlin state at filling factor $\nu=1/3$ as a simple illustration. We emphasise the Hilbert space truncation here is more general and less restrictive than the removal of unsqueezed bases in the Jack polynomial formalism\cite{jack}. The Laughlin state in the second quantised form is a Fermionic Jack polynomial with the root configuration given by $100100100\cdots1001$, where $\cdots$ represents the repeated patterns of ``$100$". All components of the Laughlin wavefunction in the occupation basis are ``squeezed" from the root configuration. An important property of the wavefunction, valid for any system size, is that a measurement of the leftmost two orbitals at the north pole will never detect more than one particle. Indeed, the \emph{reduced density matrix} of the Laughlin state restricted to the leftmost two orbitals only contain three bases: $00,10,01$; the missing basis of $11$ is because of the ``squeezing rule" of the Jack polynomial.
\begin{figure}[htb]
\includegraphics[width=\linewidth]{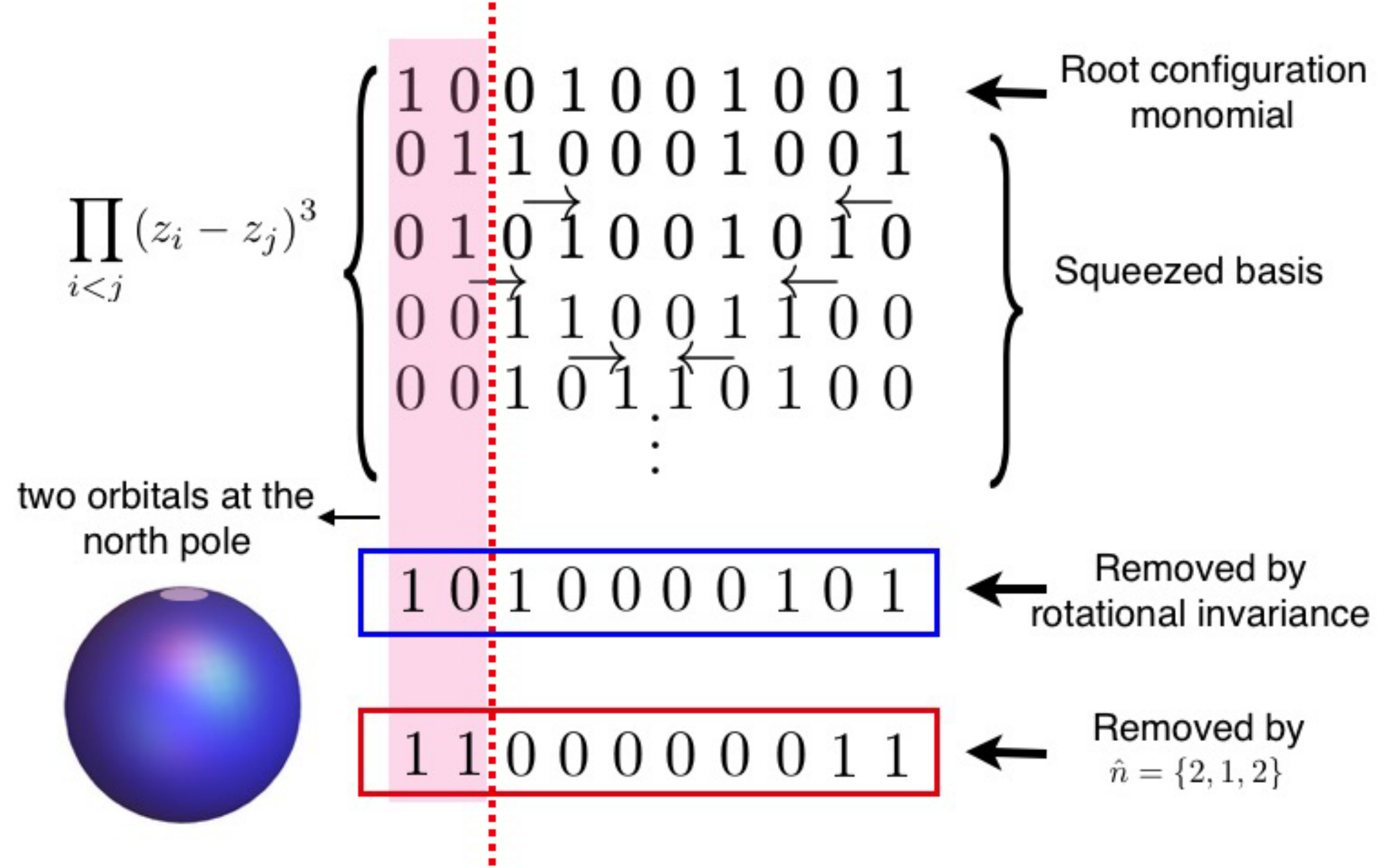}
\caption{Root configuration and squeezed bases of the Laughlin state on the sphere for four particles. The leftmost two orbitals corresponds to the north pole, and no basis has those two orbitals both occupied. Other bases with $L_z=0$ are either removed by $\mathcal C_{\hat n=\{2,1,2\}}$ or rotational invariance.}
\label{fig1}
\end{figure} 

We state here the key observation based on extensive numerical evidence: if we start with the full Hilbert space containing all bases squeezed \emph{and} unsqueezed from the root configuration (so that there is no need to select a root configuration by hand), and truncate away bases that contain two particles in the leftmost two orbitals, we are left with far more bases than the squeezed bases needed for the Laughlin state. However, when we choose $N_e$ and $N_\phi$ based on $p=1,q=3,S_e=0,S_\phi=2$ in Eq.(\ref{expression}), there is a \emph{unique} rotationally invariant $L=0$ state, obtained from diagonalising the total angular momentum operator $L^2$ on the sphere in the truncated Hilbert space: coefficients of \emph{all remaining} unsqueezed bases are forced to be zero by rotational invariance (see Fig.(\ref{fig1})).

A formal procedure can be set up for more general cases. Starting with a quantum Hall fluid on an infinite plane or a finite sphere, we define a set of LECs denoted as $\mathcal S_\mathcal C$, with each element indexed by a triplet of non-negative integers $\hat n=\{n,n_e,n_h\}$. We use $l_B$ to denote the magnetic length, so each magnetic flux occupies a finite area of $2\pi l_B^2$. A condition $\mathcal C_{\hat n}\in \mathcal S_\mathcal C$ physically dictates that for a measurement over any circular droplet containing $n$ fluxes\cite{footnote2}, no more than $n_e$ particles and $n_h$ holes can be detected. Thus $\mathcal C_{\hat n}$ acts as a constraint on the type of bases in the quantum Hall fluid that may have non-zero coefficients. If $n=n_e=n_h$, there is no constraint if the underlying particles are Fermions in a single LL. On spherical geometry, the LECs are only applied to the north or south pole, but more basis coefficients will vanish after imposing rotational invariance, effectively requiring LECs to be satisfied everywhere in the quantum fluid. The incompressibility of the FQH fluids originates from the finite energy cost of breaking such exclusion constraints, physically enforced by renormalised Coulomb interactions in various experimental conditions.

For actual implementations, we look at the Hilbert space of total z-component angular momentum $L_z=0$ on the sphere\cite{haldane1}, denoted by $\mathcal H_{N_\phi,N_e}$, indexed by the number of orbitals and electrons. We also define $\bar{\mathcal H}^{\hat n}_{N_\phi,N_e}$ to be the truncated Hilbert space from $\mathcal H_{N_\phi,N_e}$ where all bases not satisfying the constraint of $\mathcal C_{\hat n}$ at the north pole are removed, and $\bar{\mathcal N}^{\hat n}_{N_\phi,N_e}$ to be the number of $L=0$ states in $\bar{\mathcal H}^{\hat n}_{N_\phi,N_e}$. Here is one of the main statements of this work: some $\mathcal C_{\hat n}$ has a one-to-one correspondence to a combination of $[p,q,S_e,S_\phi]$ satisfying the following properties:
\begin{eqnarray}
&&N^d_{\phi}=\frac{q}{p}\left(N_e+S_e\right)-S_\phi\label{shift}\\
&&\bar{\mathcal N}^{\hat n}_{N^d_\phi,N_e}=1,\quad\bar{\mathcal N}^{\hat n}_{N_\phi<N^d_\phi,N_e}=0\label{ingap}
\end{eqnarray}
for all values of $N_e$ subject to the condition that $N_e+S_e=kp,k\ge 2$. In particular, $\hat n=\{n,n,n\}$ corresponds to the integer quantum Hall effect. This result is computationally checked for all numerically accessible system sizes\cite{sup}. The filling factors, topological shifts and particle clustering of the FQH states can all be unambiguously determined by specifying $\mathcal C_{\hat n}$ and the requirement of translational invariance. The basis expansion of the minimal model ground state can also be obtained as the unique $L=0$ ground state of $L^2$ operator in the truncated Hilbert space, and Eq.(\ref{ingap}) can be interpreted as the requirement for the state to be gapped and incompressible.

Incompressibility of the FQH phases also require gapped neutral excitations. We can show that LECs naturally forbid neutral excitations by looking at $L_z\neq 0$ sectors of different Hilbert spaces indexed by $N_\phi$ and $N_e$. Given any fixed $N_e$, for $N_\phi=N_\phi^d$ from Eq.(\ref{shift}), there are no highest weight states in any $L_z\neq 0$ sector. Thus the only highest weight state occurs in $L_z=0$ sector, which is the rotationally invariant ground state of the corresponding FQH phase, and all neutral excitations are excluded by the local classical constraints. Similarly for $N_\phi<N_\phi^d$, no highest weight states exist in any $L_z$ sector.

\begin{widetext}
\footnotesize
\begin{table}[h!]
\caption{The first row gives the triplet $\hat n$. With a single condition as the Hilbert space constraint, $S_e=0$ (see \cite{footnote}), and we use $[p,q,S_\phi]$ to represent a FQH state; L denotes the Laughlin state, Pf denotes the Pfaffian, while R denotes other states in the RR series. P denotes states from other projection Hamiltonians\cite{simon}.}
\centering
\begin{tabular}{ |c||c|c|c|c|c|c|c|c|c|c|} 
 \hline
 &$\{2,1,2\}$&$\{3,2,3\}$&$\{3,1,3\}$&$\{4,3,4\}$&$\{4,2,4\}$&$\{4,1,4\}$&$\{5,4,5\}$&$\{5,3,5\}$&$\{5,2,5\}$&$\{5,1,5\}$   \\ 
 \Xhline{3\arrayrulewidth}
$N_e=2,3,4\cdots$ &L:[1,3,2]&&L:[1,5,4]&&&L:[1,7,6]&&&&L:[1,9,8]   \\ 
 \hline
$N_e=4,6,8\cdots$ &&Pf:[2,4,2]&&&$H:[2,6,4]$&&&&P:[2,8,6]&   \\ 
 \hline
$N_e=6,9,12\cdots$  &&&&R:[3,5,2]&&&&P:[3,7,4]&&   \\ 
\hline
$N_e=8,12,14\cdots$ &&&&&&&R:[4,6,2]&&&   \\ 
\hline
\end{tabular}
\label{t1}
\end{table}
\end{widetext}
\normalsize

A few FQH states and their corresponding $\hat n$ are listed in Table.(\ref{t1}). In particular, each $\hat n={\{n,1,n\}}$ gives the usual Laughlin state at $\nu=1/\left(2n-1\right)$, and each $\hat n={\{n,n-1,n\}}$ gives the $Z_3$ parafermion (the RR series) states\cite{rr} at $\nu=\left(n-1\right)/\left(n+1\right)$. States from $\hat n={\{n,m,n\}}$ and $\hat n={\{kn,km,kn\}}$ have the same filling factor but different shifts. For example, $\hat n=\{4,2,4\}$ gives the Haffnian state\cite{simon} at $\nu=2/6, S_\phi=4$. There is also the general relationship that $\hat n={\{n,m,n\}}$ corresponds to the incompressible FQH state at $\nu=m/\left(2n-m\right)$ and $S_e=0,S_\phi=2\left(n-m\right)$. Many of these states were not studied before.

All filling factors with the corresponding shifts proposed in\cite{simon} (except for the states at filling factors $\nu=2/5,2/9$, which we will discuss later) can be included in Table.(\ref{t1}). Even in cases where projection Hamiltonians cannot uniquely determine the zero energy ground states, our scheme can lead to a unique model ground state at corresponding $\nu$ and $S_\phi$ with transparent physical interpretations. For example, using the Hamiltonian notations in Ref.\cite{simon}, the zero energy eigenstates of $P^{5}_4$ are not unique at filling factor $\nu=3/7$, yet a unique state can be obtained with $\hat n={\{5,3,5\}}$, requiring no detection of more than three particles from a measurement in a circular droplet containing five fluxes. It is also worth mentioning that every state in Table.(\ref{t1}) has its particle-hole (PH) conjugate state. They can all be uniquely determined by imposing a single condition of $\hat n={\{n,n,m\}}$. Naturally for the PH conjugate states, the condition of ``highest density" is referring to the density of the ``holes". Thus in Eq.(\ref{ingap}) we have $\bar{\mathcal N}^{\hat n}_{N_\phi>N^d_\phi,N_e}=0$ instead.

The conspicuous omissions from Table.(\ref{t1}) are the Gaffnian state (unique zero energy state of $P^{3}_3$)\cite{simon2}, the state at $\nu=2/9$ (unique zero energy state of $P^{9}_3$), and some of the filling factors where the Jain or hierarchical series are expected to occur (e.g. at $\nu=4/9$). We now proceed to show that a number of these missing states can also be realised when the $L_z=0$ Hilbert space is truncated by more than one $\mathcal C_{\hat n}$. Explicit numerical computation shows the Gaffnian state at $\nu=2/5$ with $S_e=0,S_\phi=3$ is a unique translationally invariant state when \emph{either} $\mathcal C_{\hat n_1=\{2,1,2\}}$ \emph{or} $\mathcal C_{\hat n_2=\{5,2,5\}}$ is satisfied by the $L_z=0$ bases. We denote such LECs by $\mathcal S_{\hat n_1\hat n_2}^{or}$. This implies that either \textbf{a).} a measurement of a circular droplet of $2$ fluxes can at most detect \emph{one} particle, or \textbf{b).} a measurement of a circular droplet of $5$ fluxes can at most detect \emph{two} particles. Intuitively, this is reminiscent of the hierarchical construction\cite{haldane1,halperin} or the CF picture for the $\nu=2/5$ state\cite{jain1,haldane83}. Condition \textbf{a).} is the same as the Laughlin state at $\nu=1/3$, induced effectively by a finite energy gap when $C_{\{2,1,2\}}$ is violated. However at $\nu=2/5$ the violation of \textbf{a).} is allowed as long as there is another effective energy gap for $C_{\{5,2,5\}}$. It is tempting to associate the former with the creation of quasielectrons, while the latter with the incompressibility of the quantum fluid formed by the quasielectrons themselves.

It is important to note, however, the Gaffnian state is the \emph{only} translationally invariant highest density state determined by any $\mathcal S_{\hat n_1\hat n_2}^{or}$. We discuss about its relationship with the Jain state at $\nu=2/5$ with more details in the Supplementary materials\cite{sup}. Similarly the state at $\nu=2/9$ can be determined by $\mathcal S_{\{3,1,3\}\{9,2,9\}}^{or}$, which has very high overlap with the Jain state at the same filling and shift. We have also scanned through all possible combinations of two $\mathcal C_{\hat n}$'s, and no states at $\nu=4/9$ can be uniquely determined with $\mathcal S_{\hat n_1\hat n_2}^{or}$. However, a unique state at the same filling factor but different shift can be determined by a single condition with $\{7,7,5\}$, which is the particle-hole conjugates of the $\nu=5/9$ states in Table.(\ref{t1}). Another interesting finding is that there is an incompressible state at $\nu=3/7$ and $S_\phi=4$ corresponding to $\mathcal S^{or}_{\hat n_1\hat n_2}$ with $\hat n_1=\{2,1,2\},\hat n_2=\{6,3,6\}$, which is a competing state at the same $[p,q,S_e,S_\phi]$ to the one corresponding to a single condition with $\hat n=\{5,3,5\}$ (see Table.(\ref{t1})). The overlap of these two states quickly goes to zero as we increase the system size, suggesting they belong to two different topological phases.

\begin{figure}[htb]
\includegraphics[width=9cm]{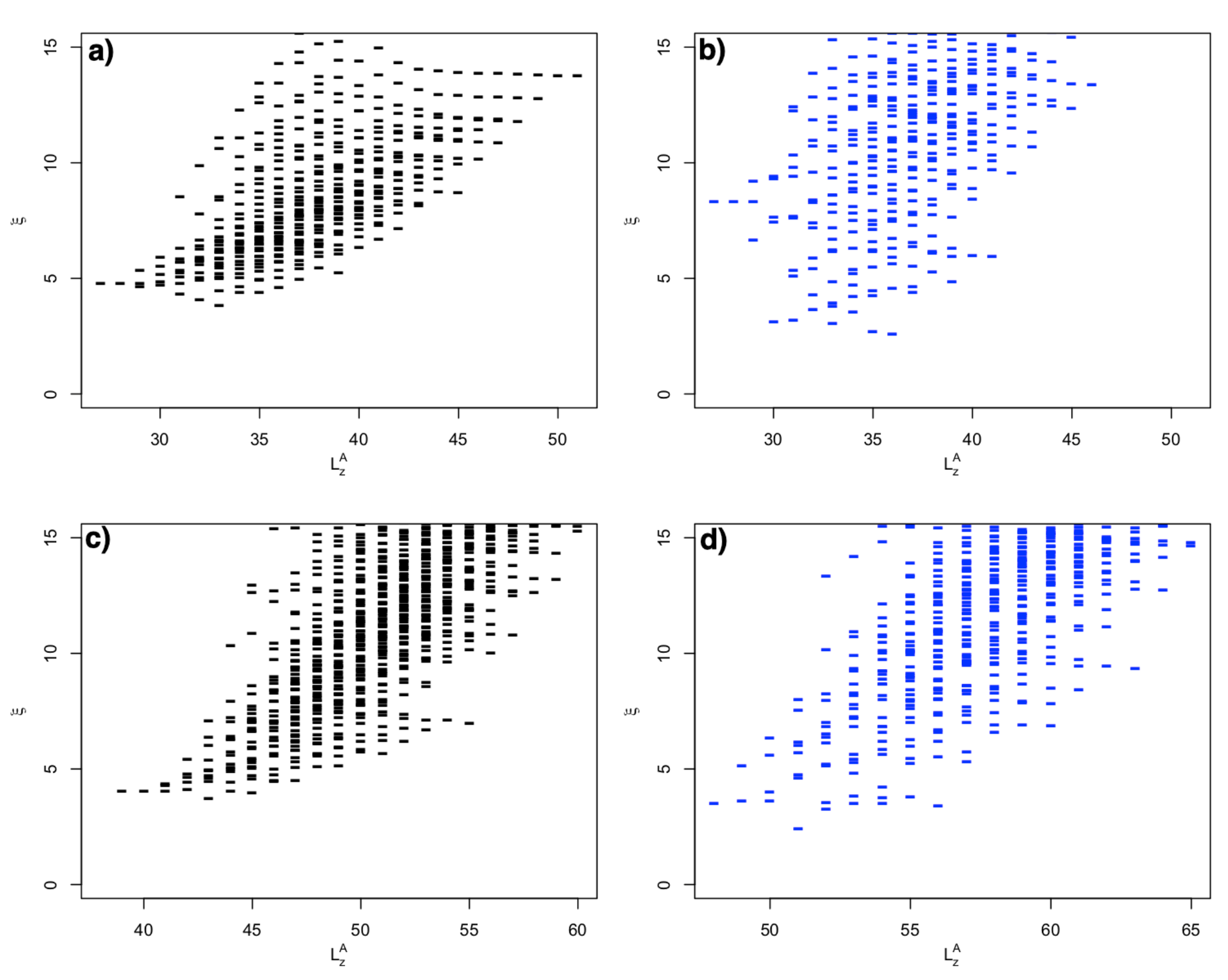}
\caption{The entanglement spectra of the $\nu=3/7$ state with $15$ particles from a). $\mathcal C_{\hat n=\{5,3,5\}}$  and b).$\mathcal S^{or}_{\hat n_1\hat n_2}$ with $\hat n_1=\{2,1,2\},\hat n_2=\{6,3,6\}$. c). $\nu=2/6$ Haffnian with 12 particles, and d). $\nu=2/5$ Gaffnian with 16 particles. The subsystem is always chosen to be the one for which the reduced density matrix has the highest dimension.}
\label{fig2}
\end{figure}

To better understand the differences between the two states at $\nu=3/7$ obtained from qualitatively different LECs, we use topological entanglement spectrum (ES)\cite{haldane3} to analyse these many-body states. One should note that our model states are explicitly constructed from a truncated Hilbert space within a single LL, thus only the topological part of ES will be present. It is worth noting from Fig.(\ref{fig2}) that even though the two $\nu=3/7$ states have an overlap of $0.02$, their ES have the exact same counting. However the state from $\mathcal S_{\hat n_1\hat n_2}^{or}$ apparently has multiple low-lying branches, while the state from $\hat n=\{5,3,5\}$ only has a single branch. Multiple low lying branches also appear from the Gaffnian state, which similarly requires two conditions to be determined uniquely. In contrast, the Haffnian state (or any other states) from a single condition only has one low-lying branch in its ES (see Fig.(\ref{fig2})). This could be another interesting physical interpretation of the roles played by $\mathcal C_{\hat n}$, which warrants further studies.

{\it Conclusions --}
We show that the incompressibility of the FQH states arises from translational invariance and constraints on the Hilbert space defined by a set of LECs denoted $\mathcal C_{\hat n}$. Thus topological properties of many FQH states can be determined classically, dictated by what can or cannot be measured from the quantum Hall fluids. Regarding the absence of Hamiltonians in this approach, one should note that the universal topological properties of the FQH phases should not depend on the Hamiltonian details. Moreover, the complexity of the model Hamiltonians does not a priori imply the corresponding FQH state is hard to realise in experiments. For example, the model Hamiltonian for the Moore-Read state is three-body, but the state can be stabilised by realistic two-body interactions, and is easier to realise in experiments than many Laughlin states with two-body model Hamiltonians.

In addition to conceptual insights, we formulated an algorithm to determine an intrinsic commensurability condition between the number of particles and number of orbitals on genus 0 geometry. The commensurability condition uniquely determines the filling factors, topological shifts and clustering properties of potential FQH fluids that can be realised in principle. The algorithm also allows efficient construction of model ground states of these FQH fluids by diagonalising a two-body operator in a truncated Hilbert space. 

It is tempting to speculate that the minimal models arising from Hilbert space truncation could be universal for all possible FQH states. In particular, it implies different FQH phases can be unambiguously classified by the LECs, and the truncation of permissible Hilbert space could be characteristic of the topological nature of the FQH states. Not only is this manifested by the missing of generic states from the entanglement spectrums of the model ground states, it is also a unifying description for both the \emph{integer and fractional} quantum Hall effect. This is because the integer quantum Hall effect is uniquely defined by translational invariance and the truncation of the bases from higher LLs.

Nevertheless, our scheme cannot account for some FQH states proposed in the literature. For example, while we can find states at $\nu=2/5,2/9\text{ and }3/7$ that have very high overlaps with the Jain states, some Jain or hierarchical states are missing; at those filling factors, incompressible states with different topological natures are obtained with our schemes. A detailed discussion of the close relationship between the LEC and composite fermion approaches will be presented elsewhere.

\begin{acknowledgments}
{\sl Acknowledgements.} I thank F.D.M. Haldane, Jainendra Jain, Ying-hai Wu and Zlatko Papic for useful discussions. This work is supported by the NTU grant for Nanyang Assistant Professorship.
\end{acknowledgments}

\pagebreak
\widetext
\begin{center}
\textbf{\large Supplemental Online Material for ``Emergent Commensurability from Hilbert Space Truncation in Fractional Quantum Hall Fluids" }\\[5pt]
\vspace{0.1cm}
\begin{quote}
{\small In this supplementary material we include some detailed examples on how the filling factors, topological shifts and particle clustering can be unambiguously determined by imposing translational invariance on the Hilbert space truncated by one or more LECs we defined in the main text. We also discuss about the Jain states in the context of the principles we proposed in the main text, which could be interesting topics for further research.}\\[20pt]
\end{quote}
\end{center}
\setcounter{equation}{0}
\setcounter{figure}{0}
\setcounter{table}{0}
\setcounter{page}{1}
\setcounter{section}{0}
\makeatletter
\renewcommand{\theequation}{S\arabic{equation}}
\renewcommand{\thefigure}{S\arabic{figure}}
\renewcommand{\thesection}{S\Roman{section}}
\renewcommand{\thepage}{S\arabic{page}}
\vspace{1cm}
\twocolumngrid

\section{S1. Examples of correspondence between $\mathcal C_{\hat n},\mathcal S^{or}_{\hat n_1\hat n_2}$ and $[p,q,S_e,S_\phi]$}

We illustrate in details how a single condition, or two conditions, can be used to unambiguously determine the filling factor $\nu=p/q$ and the topological shifts $S_e,S_\phi$ of a potential FQH states. As a first example, we take a single condition $\mathcal C_{\hat n}$ with $\hat n=\{4,3,4\}$, so that any measurement of a circular droplet of four fluxes on a translationally invariant quantum fluid will not detect more than three electrons. The task is to scan over all possible combinations of $N_e$, the number of electrons, and $N_\phi$, the number of fluxes, and to look for patterns. The empirical rule of thumb is that for $\hat n=\{n,m,n\}$, the pattern emerges when the minimal number of electrons $N_e=2m$. In this case, it starts with $N_e=6$.

It is obvious that $N_\phi\ge N_e=6$ for FQH effects. On the sphere, for each value of $N_\phi$, the $L_z=0$ sub-Hilbert space, denoted as $\mathcal H_{N_\phi,6}$, can be easily constructed. For example with $N_\phi=10$, all basis are squeezed from the dominant root configuration $1110000111$. We now start to remove from $\mathcal H_{N_\phi,6}$ all basis that contain more than three particles in the leftmost four orbitals. It turns out for $N_e=6$, no basis are removed for $N_\phi>6$, and we have $\bar{\mathcal H}^{\{4,3,4\}}_{N_\phi>6,6}=\mathcal H_{N_\phi>6,6}$. For $N_\phi=6$, the only basis given by $1111111$ is removed and $\bar{\mathcal H}^{\{4,3,4\}}_{6,6}=\emptyset$.

We now start to look for the number of rotationally invariant states by diagonalising $L^2$ in $\bar{\mathcal H}^{\{4,3,4\}}_{N_\phi,6}$, which is denoted as $\bar{\mathcal N}^{\{4,3,4\}}_{N_\phi,6}$. It turns out we have{
\begin{eqnarray}\label{six}
&&\bar{\mathcal N}^{\{4,3,4\}}_{6,6}=\bar{\mathcal N}^{\{4,3,4\}}_{7,6}=0\nonumber\\
&&\bar{\mathcal N}^{\{4,3,4\}}_{8,6}=1\nonumber\\
&&\bar{\mathcal N}^{\{4,3,4\}}_{9,6}=0\nonumber\\
&&\bar{\mathcal N}^{\{4,3,4\}}_{10,6}=2\nonumber\\
&&\bar{\mathcal N}^{\{4,3,4\}}_{11,6}=0\nonumber\\
&&\bar{\mathcal N}^{\{4,3,4\}}_{12,6}=3\nonumber\\
&&\vdots
\end{eqnarray}
In general $\bar{\mathcal N}^{\{4,3,4\}}_{N_\phi,6}$ increases with $N_\phi$, though not monotonically.

The same procedure can be done with $N_e=7,8$, we will not record the results here to avoid clutter. The interesting results are from $N_e=9$, note in this case $\bar{\mathcal H}^{\{4,3,4\}}_{N_\phi\ge,9}\in\mathcal H_{N_\phi\ge,9}$, and for each $N_\phi$, the constraint of $\mathcal C_{\{4,3,4\}}$ will truncate away some basis in the $L_z=0$ sub-Hilbert space. The number of rotationally invariant states in the truncated Hilbert space for each $N_\phi$ is given as follows:
\begin{eqnarray}\label{nine}
&& \bar{\mathcal N}^{\{4,3,4\}}_{9,9}=\bar{\mathcal N}^{\{4,3,4\}}_{10,9}=\bar{\mathcal N}^{\{4,3,4\}}_{11,9}=\bar{\mathcal N}^{\{4,3,4\}}_{12,9}=0\nonumber\\
&&\bar{\mathcal N}^{\{4,3,4\}}_{13,9}=1\nonumber\\
&&\bar{\mathcal N}^{\{4,3,4\}}_{14,9}=\bar{\mathcal N}^{\{4,3,4\}}_{15,9}=\bar{\mathcal N}^{\{4,3,4\}}_{16,9}=0\nonumber\\
&&\bar{\mathcal N}^{\{4,3,4\}}_{17,9}=6\nonumber\\
&&\bar{\mathcal N}^{\{4,3,4\}}_{18,9}=\bar{\mathcal N}^{\{4,3,4\}}_{19,9}=\bar{\mathcal N}^{\{4,3,4\}}_{20,9}=0\nonumber\\
&&\bar{\mathcal N}^{\{4,3,4\}}_{21,9}=25\nonumber\\
&&\bar{\mathcal N}^{\{4,3,4\}}_{22,9}=0\nonumber\\
&&\bar{\mathcal N}^{\{4,3,4\}}_{23,9}=25\nonumber\\
&&\vdots
\end{eqnarray}
Similarly, for $N_e=12$, we have the following results:
\begin{eqnarray}\label{twelve}
&& \bar{\mathcal N}^{\{4,3,4\}}_{12\le N_\phi\le 17,12}=0\nonumber\\
&&\bar{\mathcal N}^{\{4,3,4\}}_{18,12}=1\nonumber\\
&&\bar{\mathcal N}^{\{4,3,4\}}_{19,12}=1\nonumber\\
&&\bar{\mathcal N}^{\{4,3,4\}}_{20,12}=4\nonumber\\
&&\bar{\mathcal N}^{\{4,3,4\}}_{21,12}=6\nonumber\\
&&\bar{\mathcal N}^{\{4,3,4\}}_{22,12}=16\nonumber\\
&&\vdots
\end{eqnarray}
We can thus clearly see the pattern that for the set of values $[p,q,S_e,S_\phi]=[3,5,0,2]$, the following commensurability condition is satisfied:
\begin{eqnarray}
&&N^d_\phi=\frac{5}{3}N_e-2\\
&&\bar{\mathcal N}^{\{4,3,4\}}_{N^d_\phi,N_e}=1,\quad\bar{\mathcal N}^{\{4,3,4\}}_{N_\phi<N^d_\phi,N_e}=0\label{gap}
\end{eqnarray}
with $N_e=3k$ and $k\ge2$. Moreover, the unique rotationally invariant state in the Hilbert space of $\bar{\mathcal H}^{\{4,3,4\}}_{N_\phi^d,N_e}$ is the Read-Rezayi state at $\nu=3/5$, a Fermionic Jack polynomial with root configuration of $1110011100\cdots 11100111$, or the famous Fibonacci state. We have numerically checked such to be the case up to 18 particles.

As another example, we look for $[p,q,S_e,S_\phi]$ corresponding to $\mathcal S^{or}_{\hat n_1,\hat n_2}$, with $\hat n_1=\{2,1,2\},\hat n_2=\{6,3,6\}$. Using $N_e=9,N_\phi=17$ as an example, the entire $L_z=0$ sub-Hilbert space is squeezed from $11110000100001111$; among all the squeezed basis, if the leftmost two orbitals contain zero or one particles (e.g. $00001111110000$ or $10000111100001$), these basis will be kept. If the leftmost two orbitals contain two particles, but the leftmost six orbitals contain no more than three particles (e.g. $111000011100000111$), such basis will also be kept. If both conditions are violated (e.g. $11110000100001111$), such basis will be truncated. For notational convenience we denoted truncated Hilbert space as $\bar{\mathcal H}_{N_\phi,N_e}^{\hat n_1\hat n_2}$, and the number of $L=0$ states in this Hilbert space as $\bar{\mathcal N}_{N_\phi,N_e}^{\hat n_1\hat n_2}$.

With $\hat n_1=\{2,1,2\},\hat n_2=\{6,3,6\}$, the pattern again starts with $N_e=6,9,12,\cdots$. At $N_e=6$, the imposition of $\mathcal S^{or}_{\hat n_1,\hat n_2}$ again does not remove any basis except for $N_\phi=N_e=6$. Thus identical to Eq.(\ref{six}) we have the following:
\begin{eqnarray}\label{sixa}
&&\bar{ \mathcal N}^{\hat n_1\hat n_2}_{6,6}=\bar{\mathcal N}^{\hat n_1\hat n_2}_{7,6}=\bar{\mathcal N}^{\hat n_1\hat n_2}_{8,6}=\bar{\mathcal N}^{\hat n_1\hat n_2}_{9,6}=0\nonumber\\
&&\bar{\mathcal N}^{\hat n_1\hat n_2}_{10,6}=1\nonumber\\
&&\bar{\mathcal N}^{\hat n_1\hat n_2}_{11,6}=0\nonumber\\
&&\bar{\mathcal N}^{\hat n_1\hat n_2}_{12,6}=3\nonumber\\
&&\bar{\mathcal N}^{\hat n_1\hat n_2}_{13,6}=0\nonumber\\
&&\bar{\mathcal N}^{\hat n_1\hat n_2}_{14,6}=4\nonumber\\
&&\vdots
\end{eqnarray}
With $N_e=9$ we have the following:
\begin{eqnarray}\label{ninea}
&& \bar{\mathcal N}^{\hat n_1\hat n_2}_{9\le N_\phi\le 16,9}=0\nonumber\\
&&\bar{\mathcal N}^{\hat n_1\hat n_2}_{17,9}=1\nonumber\\
&&\bar{\mathcal N}^{\hat n_1\hat n_2}_{18,9}=\bar{\mathcal N}^{\hat n_1\hat n_2}_{19,9}=\bar{\mathcal N}^{\hat n_1\hat n_2}_{20,9}=0\nonumber\\
&&\bar{\mathcal N}^{\hat n_1\hat n_2}_{21,9}=14\nonumber\\
&&\bar{\mathcal N}^{\hat n_1\hat n_2}_{22,9}=0\nonumber\\
&&\bar{\mathcal N}^{\hat n_1\hat n_2}_{23,9}=15\nonumber\\
&&\vdots
\end{eqnarray}
And with $N_e=12$ we have the following:
\begin{eqnarray}\label{twelvea}
&& \bar{\mathcal N}^{\hat n_1\hat n_2}_{12\le N_\phi\le 23,12}=0\nonumber\\
&&\bar{\mathcal N}^{\hat n_1\hat n_2}_{24,12}=1\nonumber\\
&&\bar{\mathcal N}^{\hat n_1\hat n_2}_{25,12}=1\nonumber\\
&&\bar{\mathcal N}^{\hat n_1\hat n_2}_{26,12}=9\nonumber\\
&&\bar{\mathcal N}^{\hat n_1\hat n_2}_{21,12}=20\nonumber\\
&&\vdots
\end{eqnarray}
Thus for $\mathcal S^{or}_{\hat n_1,\hat n_2}$ with $\hat n_1=\{2,1,2\},\hat n_2=\{6,3,6\}$, the corresponding commensurability condition is as follows:
\begin{eqnarray}
&&N^d_\phi=\frac{7}{3}N_e-4\\
&&\bar{\mathcal N}^{\hat n_1\hat n_2}_{N^d_\phi,N_e}=1,\quad\bar{\mathcal N}^{\hat n_1\hat n_2}_{N_\phi<N^d_\phi,N_e}=0\label{gap}
\end{eqnarray}
with $N_e=3k$ and $k\ge2$. The unique rotationally invariant state in the Hilbert space of $\bar{\mathcal H}^{\hat n_1\hat n_2}_{N_\phi^d,N_e}$ is obtained up to $N_e=15$. This is the minimal model state for the Jain state at $\nu=3/7$.

In both two examples above, we have $p=3$ so the model states only exist when $N_e$ is divisible by $3$, indicating its non-Abelian or multicomponent nature. For the Laughlin states, we have $p=1$, and the Moore Read state gives $p=2$, as can be obtained with the same procedure described in this section.

\section{S2. The Jain States from Composite Fermion Picture}

In this section we give a brief discussion on the interesting questions arising from the compatibility and incongruity between the commensurability conditions and the composite fermion construction. A more detailed analysis will be presented elsewhere. The simplest Jain state occurs at $[p,q,S_e,S_\phi]=[2,5,0,3]$ with $\nu=2/5$ and topological shift $S_\phi=3$. In the paradigm of composite fermions, the state can be interpreted as the integer quantum Hall effect of composite fermions made of one electron and two fluxes, when the lowest two ``CF levels" are completely filled by these composite fermions.

The trial wavefunction of the Jain state at $\nu=2/5$ is obtained by projecting the integer quantum Hall wavefunction of composite fermions (containing basis in higher LLs) into the lowest Landau level. It is also worth pointing out that the $\nu=2/5$ state with $S_\phi=3$ can also be understood as a hierarchical state, where the quasielelctrons of the Laughlin state at $\nu=1/3$ form its own incompressible state. In both pictures, the state is spin polarised, Abelian and multicomponent. The two phenomenological pictures are physically distinct, yet the trial wavefunctions obtained seems to represent the same topological phase, possibly suggesting more fundamental elements not reflected by these two phenomenological pictures.

In our commensurability scheme, no single $\mathcal C_{\hat n}$ leads to the commensurability condition of $[p,q,S_e,S_\phi]=[2,5,0,3]$. We have also scanned over the combination of two or three conditions, and only $\mathcal S^{or}_{\hat n_1,\hat n_2}$ with $\hat n_1=\{2,1,2\},\hat n_2=\{5,2,5\}$ leads to $[p,q,S_e,S_\phi]=[2,5,0,3]$, a unique state with the physical interpretation that either a circular droplet of two fluxes contains no more than one particle (the condition that corresponds to the $\nu=1/3$ Laughlin state), or a circular droplet of five fluxes contains no more than two particles. This physical interpretation is intuitively very much compatible with the hierarchical picture for the $\nu=2/5$ state.

What is interesting is that the model state from $\mathcal S^{or}_{\hat n_1,\hat n_2}$ with $\hat n_1=\{2,1,2\},\hat n_2=\{5,2,5\}$ is actually the Gaffnian state, which is apparently the only model state for $[p,q,S_e,S_\phi]=[2,5,0,3]$ based on the principles we proposed in the main text. This raises the intriguing question of the relationship between the Jain state and the Gaffnian state at $\nu=2/5$. The two states have very high overlap, and similar features in their entanglement spectrum. However it is also believed they are topologically distinct\cite{simon2}. The CF picture dictates the Jain state to be Abelian, while the Gaffnian and its quasihole states were originally constructed from CFT correlators, implying it is non-Abelian. Moreover, it has also been argued that the Gaffnian state should be gapless\cite{simon2,bernevig, bernevig2}.

We argue here that the Gaffnian state \emph{could be} the minimal model state of the Jain state or the hierarchical state at $\nu=2/5,S_\phi=3$. This is not contradictory to the claims about the Gaffnian state from the CFT picture, as the detailed analysis in \cite{bernevig} shows from topological entanglement entropy that the quasiparticle excitations of the Gaffnian state is \emph{Abelian}, and the bulk correlation length only seems to go to zero in the thermodynamic limit for the non-Abelian sector: the Abelian vacuum sector remains finite. The subtle relationship between Gaffnian and Jain state was also reported in Ref.\cite{jain3} from the behaviours of quasiparticle excitations. The claim of gapless-ness of the Gaffnian from the CFT perspective only indicates the state happens to be the zero energy ground state of a certain microscopic projection Hamiltonian that is gapless. It does not forbid the state from capturing the essential topological properties of the ground state of a more realistic, incompressible Hamiltonian.

Indeed, in many perspectives the principles we proposed in the main text are quite general (e.g. it works for the Jain state at $\nu=3/7$), and the Gaffnian state naturally emerges if the realistic microscopic Hamiltonian gives a significant energy punishment (as compared to other energy scales) when three or more particles appear in a five-flux droplet, \emph{and} two particles appear in a two-flux droplet within this five-flux droplet, which is no more special than any other FQH states mentioned in this work. This also explains the high overlap between the Gaffnian and the Jain state, and seems to indicate certain limitations of the CF construction and the CFT perspective in the context of the fractional quantum Hall physics.


\begin{thebibliography}{99}
\bibitem{prange} D. C. Tsui, H. L. Stormer, and A. C. Gossard, Phys. Rev. Lett. {\bf 48}, 1559 (1982). R. Prange and S. Girvin, The Quantum Hall effect, Graduate texts in contemporary physics (Springer- Verlag, 1987), ISBN 9783540962861
\bibitem{laughlin} R.B. Laughlin, Phys. Rev. Lett. {\bf 50}, 1395 (1983).
\bibitem{sarmabook} Perspectives in Quantum Hall Effects, edited by S. Das Sarma and A. Pinczuk (Wiley, New York, 1996).
\bibitem{mr} G. Moore and N. Read, Nucl. Phys. B. {\bf 360}, 362 (1991).
\bibitem{rr} N. Read and E. Rezayi, Phys.Rev. B. {\bf 59}, 8084 (1999).
\bibitem{haldane1}F.D.M. Haldane, Phys. Rev. Lett. {\bf 51}, 605 (1983).
\bibitem{jain1} J.K. Jain, Phys. Rev. Lett. {\bf 63}, 199 (1989).
\bibitem{haldane83} F.D.M. Haldane, Phys. Rev. Lett. {\bf 51}, 605 (1983).
\bibitem{jain2} J.K. Jain, Phys. Rev. B. {\bf 40}, 8079 (1989).
\bibitem{wen1} X.G. Wen, Phys. Rev. Lett. {\bf 66}, 802 (1991).
\bibitem{jainbook} Composite Fermions: A Unified View of the Quantum Hall Regime, edited by Olle Heinonen (World Scientific, New York, 1998).
\bibitem{jack} B.A. Bernevig and F.D.M. Haldane, Phys. Rev. Lett. {\bf 100}, 246802 (2008).
\bibitem{bernevig}N. Regnault, B.A. Bernevig and F.D.M. Haldane, Phys. Rev. Lett. {\bf 103}, 016801(2009).
\bibitem{yang1} Bo Yang, Z-X. Hu, Z. Papic and F.D.M. Haldane, Phys. Rev. Lett. {\bf 108}, 256807 (2012).
\bibitem{yang2} Bo Yang and F.D.M. Haldane, Phys. Rev. Lett. {\bf 112}, 026804 (2014).
\bibitem{simon} S.H. Simon, E.H. Rezayi and N.R. Cooper, Phys. Rev. B. {\bf 75}, 075318 (2007).
\bibitem{wen2} X.G. Wen and A. Zee, Phys. Rev. B. {\bf 46}, 2290 (1992).
\bibitem{footnote} In this paper we only look at cases with $S_e=0$, but there are also examples of nonzero $S_e$ shown by Thomale et.al. (unpublished results).
\bibitem{sreejith} G.J. Sreejith, M. Fremling, G.S. Jeon and J.K. Jain, arXiv: 1809.06325.
\bibitem{haldane2} F.D.M. Haldane, Phys. Rev. Lett. {\bf 67}, 937 (1991).
\bibitem{oshikawa} M. Oshikawa, Phys. Rev. Lett. {\bf 84}, 1535 (2000).
\bibitem{footnote1} In principle we can also have symmetry protected ground state degeneracy gapped from the rest of the spectrum, or a manifold of ground state degeneracy with the dimension growing sub-extensively with the system size. In this work we focus on non-degenerate ground state and we will defer such FQH topological phases to future works.
\bibitem{haldane3}H. Li and F.D.M. Haldane, Phys. Rev. Lett. {\bf 101}, 010504 (2008).
\bibitem{simon2} S.H. Simon, E.H. Rezayi, N.R. Cooper and I. Berdnikov, Phys. Rev. B. {\bf 75}, 075317 (2007).
\bibitem{halperin}B.I. Halperin, Phys. Rev. Lett. {\bf 52}, 1583 (1984).
\bibitem{bernevig2} B. Estienne, N. Regnault and B.A. Bernevig, Phys. Rev. Lett. {\bf 114}, 186801 (2015).
\bibitem{jain3} C. Toke and J.K. Jain, Phys. Rev. B. {\bf 80}, 205301 (2009).
\bibitem{footnote2} Such circular droplets are not exactly defined in real space, but rather in a single LL with an orbital cut.
\bibitem{sup} See Supplemental Material at [URL will be inserted by publisher for] for some specific examples of the LEC implementation, and for some further discussions on the Gaffnian state and the Jain state at $\nu=2/5$.
\end{thebibliography}

\begin{thebibliography}{99}
\bibitem{bernevig}N. Regnault, B.A. Bernevig and F.D.M. Haldane, Phys. Rev. Lett. {\bf 103}, 016801(2009).
\bibitem{simon2} S.H. Simon, E.H. Rezayi, N.R. Cooper and I. Berdnikov, Phys. Rev. B. {\bf 75}, 075317 (2007).
\bibitem{bernevig2} B. Estienne, N. Regnault and B.A. Bernevig, Phys. Rev. Lett. {\bf 114}, 186801 (2015).
\bibitem{jain3} C. Toke and J.K. Jain, Phys. Rev. B. {\bf 80}, 205301 (2009).\end{thebibliography}
\end{document}